\documentclass[preprint,superscriptaddress,showpacs,preprintnumbers,amsmath,amssymb,aps,pre]{revtex4-1}
\usepackage{epsfig}
\usepackage{natbib}
\usepackage{amsmath}
\usepackage{times}
\usepackage{subfigure}
\usepackage{graphicx}
\usepackage{color}
\usepackage{dsfont}

\begin{document}

\title{Uncertainty dimension and basin entropy in relativistic chaotic scattering}

\author{Juan D. Bernal}
\email[]{juandiego.bernal@urjc.es}
\affiliation{Nonlinear Dynamics, Chaos and Complex Systems Group, Departamento de
F\'{i}sica, Universidad Rey Juan Carlos, Tulip\'{a}n s/n, 28933 M\'{o}stoles, Madrid, Spain}

\author{Jes\'{u}s M. Seoane}
\affiliation{Nonlinear Dynamics, Chaos and Complex Systems Group, Departamento de
F\'{i}sica, Universidad Rey Juan Carlos, Tulip\'{a}n s/n, 28933 M\'{o}stoles, Madrid, Spain}

\author{Miguel A.F. Sanju\'{a}n}
\affiliation{Nonlinear Dynamics, Chaos and Complex Systems Group,
Departamento de F\'{i}sica, Universidad Rey Juan Carlos, Tulip\'{a}n
s/n, 28933 M\'{o}stoles, Madrid, Spain}
\affiliation{Department of Applied Informatics, Kaunas University of Technology, Studentu 50-415, Kaunas LT-51368, Lithuania}
\affiliation{Institute for Physical Science and Technology, University of Maryland, College
Park, Maryland 20742, USA}

\date{\today}

\begin{abstract}
Chaotic scattering is an important topic in nonlinear dynamics and chaos with applications in several
fields in physics and engineering. The study of this phenomenon in relativistic systems has received
little attention as compared to the Newtonian case. Here, we focus our work on the study of some relevant characteristics of the exit basin topology in the relativistic H\'{e}non-Heiles system: the uncertainty dimension, the Wada property and the basin entropy. Our main findings  for the uncertainty dimension show two different behaviors insofar we change the relativistic parameter
$\beta$, in which a crossover behavior is uncovered. This crossover point is related with the disappearance
of KAM islands in phase space that happens for velocity values above the ultra-relativistic limit, $v>0.1c$.
This result is supported by numerical simulations and also by qualitative analysis, which are in good agreement.
On the other hand, the computation of the exit basins in the phase space suggests the existence of Wada
basins for a range of $\beta<0.625$. We also studied the evolution of the exit basins in a quantitative manner by computing the basin entropy, which shows a maximum value for $\beta \approx 0.2$. This last quantity is related to the uncertainty in the prediction of the final fate of the system. Finally, our work is relevant in galactic dynamics and it also has important implications in other topics in physics as in the St\"{o}rmer problem, among others.

\end{abstract}

\maketitle

\section{Introduction} \label{sec:Introduction}

\textit{Chaotic scattering} in open Hamiltonian systems is a field broadly studied in nonlinear dynamics and chaos during the last decades. It has multiple applications in physics as, for example, the modeling of the dynamics of ions moving in electromagnetic traps, the interaction between the magnetic tail of the Earth and the solar wind, or the analysis of the escape of stars from galaxies (see Refs.~\cite{New Developments,OTT}). Chaotic scattering is defined by the interaction between an incident particle and a potential region or massive object that scatters it. This interaction occurs in such a way that the equations of motion of the test particle are nonlinear and the resultant dynamics can be chaotic. Therefore, slightly initial conditions may describe completely different trajectories and, eventually, yield to diverse final destinations. The scatterer system can be modeled by a potential, and the region where the particle is affected by that potential is called \textit{scattering region}. The influence of that potential on the particle is negligible outside the scattering region and then the motions of the incident particle are uniform and essentially free. Since the system is open, the scattering region possesses exits from which the particles may enter or escape. Quite often, particles escape from the scattering region after bouncing back and forth there for a while. In this sense, chaotic scattering is regarded as a physical manifestation of transient chaos ~\cite{Yorke_Physica,Directions_in_Chaos}.\\

On the other hand, when the velocity of an object is low, compared to the speed of light, the Newtonian approximation can be used for modeling the dynamics of that object. This is indeed the most widely accepted convention in physics and engineering calculations \cite{OHANIAN}. However, if the dynamics of the system is chaotic, the trajectories predicted by the Newtonian scheme rapidly disagree with the ones described by the special relativity theory (see Refs.~\cite{Lan, solitons, Borondo, bouncingball}). Moreover, it has been recently demonstrated that there are relevant global properties of the chaotic scattering systems that actually do depend on the effect of the Lorentz transformations ~\cite{JD relativistic}. To our knowledge, we think there are no previous research characterizing the exit basin topology of open relativistic Hamiltonian systems. For that reason, in the present work we focus our efforts on the study of some key properties that characterizes the exit basin topology of the relativistic H\'{e}non-Heiles system: the uncertainty dimension, the Wada property and the basin entropy. This is important since the exit basin topology is useful to achieve a priori very rich global insights. For example, once we know the exit basin topology of a system, we can infer the degree of unpredictability of the final state of the system by just knowing some approximate information about the initial conditions.\\

This paper is organized as follows. In Sec.~\ref{sec:Model Description}, we describe the model, \textit{i.e.}, the relativistic H\'{e}non-Heiles system. The effects of the Lorentz transformation on the uncertainty dimension of a typical scattering function are carried out in Sec.~\ref{sec:fractal dimension}. A qualitative description of the effects derived from the special relativity in the exit basins is shown in Sec.~\ref{sec:Escape Basins}. In Sec. ~\ref{sec:Basin Entropy}, we use the basin entropy, which is a novel tool developed to quantify the uncertainty of the system based on the study of the exit basin topology. Then we understand the different sources of uncertainty in the relativistic H\'{e}non-Heiles system caused by the Lorentz transformations. Finally, in Sec.~\ref{chap: conclusions} we conclude with a discussion of the main results of this work.

For the sake of clarity, throughout this paper, we refer to ``relativistic" to any effect where the Lorentz transformations have been considered. Likewise, we say that any property or object is ``nonrelativistic" or ``Newtonian" when we do not take into consideration the Lorentz transformations, but the Galilean ones.


\section{Model description} \label{sec:Model Description}
We focus our study on the effects of the relativistic corrections in a paradigmatic chaotic scattering system, the H\'{e}non-Heiles Hamiltonian. The two-dimensional potential of the H\'{e}non-Heiles system is defined by

\begin{equation}\label{eq:HH potential}
V(x,y) = \frac{1}{2}k(x^2+y^2) + \lambda (x^2y - \frac{1}{3}y^3).
\end{equation}

\begin{figure}[h]
\centering
\includegraphics[width=0.65\textwidth]{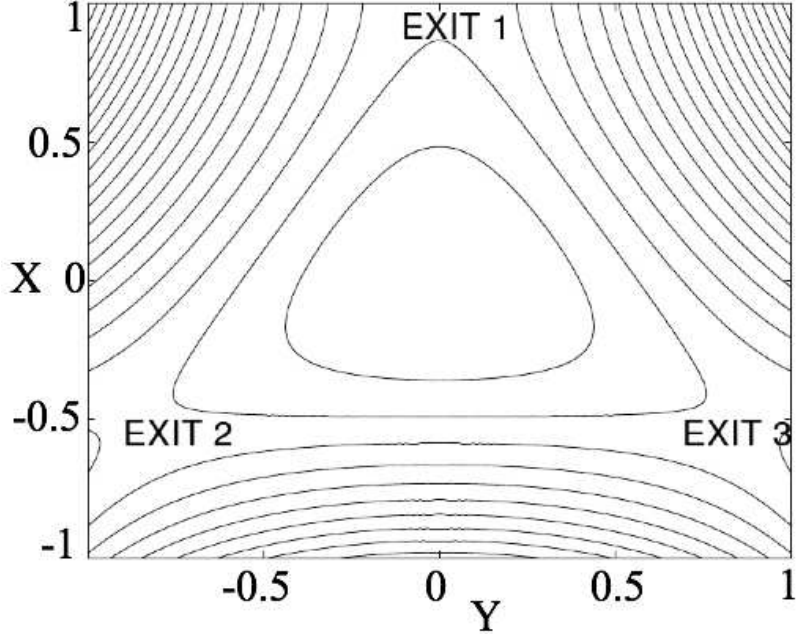}
\caption{\textbf{Isopotential curves for the H\'{e}non-Heiles potential.} They are closed for energies below the nonrelativistic threshold energy escape $E_e=1/6$. It shows three different exits for energy values above $E_e=1/6$.}
\label{Fig1}
\end{figure}

We consider the H\'{e}non-Heiles potential in a system of units where $k=\lambda=1$, see Ref.~\cite{Henon_Heiles}. Its isopotential curves can be seen in Fig.~\ref{Fig1}. Due to the triangular symmetry of the system, the exits are separated by an angle of $2\pi/3$ radians. We call Exit $1$ the upper exit ($y \rightarrow +\infty$), Exit $2$, the left one ($y\rightarrow -\infty, x\rightarrow -\infty$), and, Exit $3$, the right exit ($y\rightarrow -\infty, x\rightarrow +\infty$).

We define the nonrelativistic total mechanical energy and we call it \textit{Newtonian energy}, $E_N$, as $E_N = T(\mathbf{p}) + V(\mathbf{r})$, where $T$ is the kinetic energy of the particle, $T = \mathbf{p}^2/2m$, $\mathbf{p}$ is its linear momentum vector, $V(\mathbf{r})$ is the potential energy, and $\mathbf{r}$ it its vector position. If $E_N\in [0,1/6]$, the trajectory of any incident particle is trapped into the scattering region. For $E_N > 1/6$, the particles may eventually escape up to infinity. There are indeed three different regimes of motion depending on the initial value of the energy: (a) \textit{closed-nonhyperbolic} $E_N \in [0,1/6]$, (b) \textit{open-nonhyperbolic} $E_N \in (1/6,2/9)$ and (c) \textit{open-hyperbolic} $E_N \in [2/9,+\infty)$ \cite{022blesa}. In the first energy range, all the trajectories are trapped and there is no exit by which any particle may escape. When $E_N \in (1/6,2/9)$, the energy is large enough to allow escapes from the scattering region and the coexistence of stable invariant tori with chaotic saddles, which typically results in an algebraic decay in the survival probability of a particle in the scattering region. On the contrary, when $E_N \in [2/9,+\infty)$, the regime is open-hyperbolic and all the periodic trajectories are unstable; there is no KAM tori in phase space.

If we consider the motion of a relativistic particle moving in an external potential energy \textit{V(\textbf{r})}, the Hamiltonian (or the total energy) is:

\begin{equation}\label{eq:Hamiltoniano relativista}
H = E = \gamma mc^2 + V(\mathbf{r}) = \sqrt{m^{2}c^{4}+c^{2}\mathbf{p}^{2}}+V(\mathbf{r}),
\end{equation}

\noindent where \textit{m} is the particle's rest mass, \textit{c} is the speed of light and $\gamma$ is the Lorentz factor which is defined as:
\begin{equation}\label{eq:Lorentz factor}
\gamma = \sqrt{1 + \frac{\mathbf{p}^2}{m^2c^2}} = \frac{1}{\sqrt{1 - \frac{\mathbf{v}^2}{c^2}}}.
\end{equation}

Therefore, the Hamilton's canonical equations are:
\begin{equation}\label{eq:ecuaciones de Hamilton}
\begin{aligned}
\dot{\mathbf{p}} & = - \frac{\partial H}{\partial \mathbf{r}} = -\nabla V(\mathbf{r}),\\
\dot{\mathbf{r}} = \mathbf{v} & = \frac{\partial H}{\partial \mathbf{p}} = \frac{\mathbf{p}}{m\gamma}.
\end{aligned}
\end{equation}

When $\gamma = 1$ the Newtonian equations of motion are recovered from Eq.~\ref{eq:ecuaciones de Hamilton}. We define $\beta$ as the ratio $v/c$, where $v$ is the modulus of the vector velocity $\mathbf{v}$. Then the Lorentz factor can be rewritten as $\gamma = \frac{1}{\sqrt{1 - \beta^2}}$. Whereas $\gamma\in[1,+\infty)$, the range of values for $\beta$ is $[0,1]$. However, $\gamma$ and $\beta$ express essentially the same: how large is the velocity of the object as compared to the speed of light. From now on, we will use $\beta$ instead of $\gamma$ to show our results for mere convenience.

Taking into consideration Eqs.~\ref{eq:HH potential} and \ref{eq:ecuaciones de Hamilton}, the relativistic equations of motion of a scattering particle of unit rest mass $(m=1)$ interacting with the H\'{e}non-Heiles potential are:

\begin{equation} \label{eq: relativistic eq motion}
\begin{aligned}
\dot{x} & = \frac{p}{\gamma},\\
\dot{y} & = \frac{q}{\gamma},\\
\dot{p} & = -x - 2xy,\\
\dot{q} & = -y - x^2 + y^2,
\end{aligned}
\end{equation}

\noindent where \textit{p} and \textit{q} are the two components of the linear momentum vector \textit{\textbf{p}}.\\

\begin{figure}[htp]
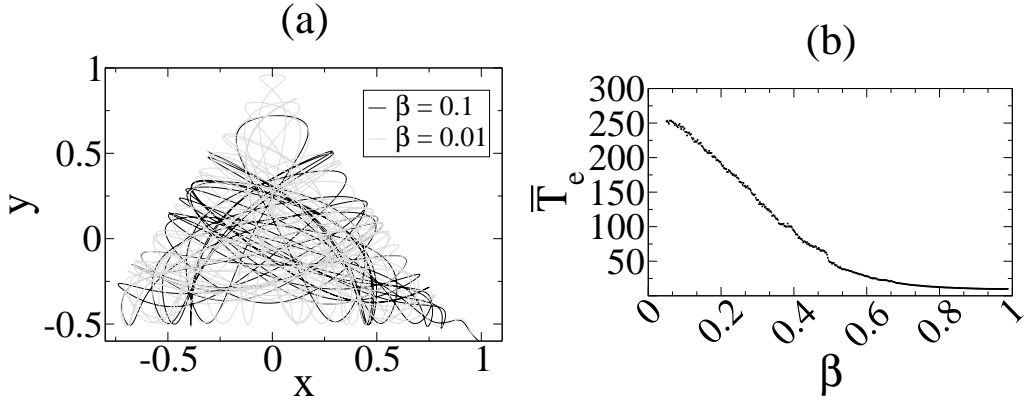

\centering
\includegraphics[width=0.4\textwidth,clip]{Fig2a.eps}\hspace{.25 cm}
\includegraphics[width=0.4\textwidth,clip]{Fig2b.eps}

\caption{(Color online) \textbf{Some previous results on relativistic chaotic scattering}. (a) The figure shows different trajectories of the same relativistic particle when it is shot at the same initial Newtonian velocity, $v = 0.583$, from the same initial conditions: $x = 0$, $y = 0$ and shooting angle $\phi = 0.1\pi$. That velocity is measured in different systems of units, representing different values of the parameter $\beta$. The black curve represents the trajectory of the particle for $\beta = 0.01$ and, in light gray, the one for $\beta = 0.1$. (b) The figure shows the average escape time, $\bar{T_e}$, of $10,000$ particles inside the scattering region with initial velocity $v = 0.583$. The initial conditions are $(x_0,y_0,\dot{x_0},\dot{y_0})= (0,0,v\cos\phi,v\sin\phi)$, with shooting angle, $\phi \in \left[0,2\pi\right]$. We use $500$ different values of $\beta$ in our calculations. There is a linear decrease of $\bar{T_e}$ up to $\beta \approx 0.4$. Indeed at $\beta \approx 0.4$ there is a leap where the linear decrease trend of $\bar{T_e}$ changes abruptly due to the KAM islands destruction.}
\label{Fig2}
\end{figure}

In the present work, we aim to isolate the effects of the variation of the Lorentz factor $\gamma$ (or $\beta$ as previously shown) from the rest of variables of the system, \textit{i.e.}, the initial velocity of the particles, its energy, etc. For this reason, during our numerical computations we have used a different system of units so that $\gamma$ be the only varying parameter in the equations of motion (Eq.~\ref{eq: relativistic eq motion}). Therefore, we analyze the evolution of the properties of the system when $\beta$ varies, by comparing these properties with the characteristics of the nonrelativistic system. Then we have to choose the same value of the initial velocity, $v = 0.583$, in different systems of units. This initial velocity corresponds to a Newtonian energy $E_N=0.17$, which is in the open-nonhyperbolic regime and quite close to the limit $E_e$. For the sake of clarity, we consider an incident particle coming from the infinity to the scattering region. Then, imagine that we measure the properties of the incident particle in the system of Planck units. In this system, the speed of light is $c = 1\;c$, that is, the unit of the variable speed is measured as a multiple of the speed of light $c$ instead of, for instance, in $m/s$. Now, suppose that, according to our measures, the rest mass of the particle is $m = 1\;m_P$ (in the Planck units, the mass is expressed as a multiple of the Planck mass $m_P$, which is $m_P \approx 2.2\times10^{-8}\;kg$). Likewise, the speed of the particle is $v = 0.583\;c$. According to the Newtonian scheme, the classical energy of the particle is $E_N = \frac{1}{2}v^2\approx0.17\;E_{P}$, where $E_P$ is the Planck energy, which is the energy unit in the Planck units ($E_{P}\approx1.96\times10^{9}\;J$). Now, we consider another incident particle with different rest mass and velocity, however we choose the International System of Units (\textit{SI}) to measure its properties. In this case, we obtain that its rest mass is again the unity, although now the rest mass is one kilogram, $m = 1\;kg$, and its velocity is $v = 0.583\;m/s$.  The speed of light in \textit{SI} is $c\approx 3\times10^{8}\;m/s$. The initial Newtonian energy of the particle is $E_N\approx0.17\;J$ in \textit{SI}, but now the initial velocity is almost negligible as compared to the speed of light, that is, $\beta = v/c = 0.583/3\times10^{8} \approx 2\times10^{-9}$. Therefore, from the perspective of the equations of motion of both particles, when we just consider the Galilean transformations, we can conclude that the behavior of the particles will be the same since $V(x,y)$ of Eq.~\ref{eq:HH potential} is equal in both cases, as long as the parameters $k=\lambda=1$ in their respective system of units. However, they are completely different when the relativistic corrections are considered because the Lorentz factor $\gamma$ affects the equations of motion by the variation of $\beta$, regardless the chosen system of units. To summarize, the objective of our numerical computations and analysis is to study the effect of $\gamma$ in the equations of motion, so the key point is to set the speed of light \textit{c} as the threshold value of the speed of the particles, regardless the system of units we may be considering. In order to give a visual example, in Fig.~\ref{Fig2}(a), we represent two different trajectories of the same relativistic particle when it is shot at the same initial Newtonian velocity, $v = 0.583$, from the same initial conditions, $x = 0$, $y = 0$ and equal initial shooting angle $\phi = 0.1\pi$. However, that velocity, measured in different systems of units, represents different values of the parameter $\beta$. The black curve is the trajectory of the particle for $\beta = 0.01$ and, in light gray, we represent the trajectory for $\beta = 0.1$. Both trajectories leave the scattering regions by the same exit, although the paths are completely different and the time spent in this region too.\\

Apart from the dependence of single trajectories on the variation of $\beta$, it is important to note that some global properties of the system, as the average escape time of the particles and the decay law of the system, do depend on the value of $\beta$ too. For example, in Fig.~\ref{Fig2}(b) we represent the average escape time $\bar{T_e}$ of a large number of incident particles \textit{vs.} $\beta$. There is a linear decrease of $\bar{T_e}$ up to $\beta \approx 0.4$. Then, there is a leap where the linear decrease trend of $\bar{T_e}$ changes abruptly. That crossover point at $\beta \approx 0.4$ is causeded by the destruction of the KAM island. This is also the point where the system dynamics changes: for $\beta \leq 0.4$ the system dynamics is nonhyperbolic and the decay law of the particles is algebraic. Then the fraction $P(t)$ of the particles remaining in the scattering region obeys the law $P(t) \sim t^{-z}$. However, for $\beta > 0.4$, the regime is hyperbolic and the decay law is exponential: $P(t) \sim e^{-\tau t}$. The exponent $z$ of the algebraic decay law and the parameter $\beta$ are both related in a quadractic manner, such that $z(\beta) = Z_0 + Z_1\beta + Z_2\beta^2$. Likewise, the relation between the parameter $\tau$ of the hyperbolic regime and the relativistic parameter $\beta$ follows a monotonically increasing quadratic relation as $\tau(\beta) \approx \Gamma_0 + \Gamma_1\beta + \Gamma_2\beta^2$, for $\beta \in (0.4,0.95]$. Note that the maximun value of $z$ in the algebraic regime is when the KAM structures are destroyed at $\beta \approx 0.4$. The value of $z$ at this point is $z \approx 1.5$ (for more details see ~\cite{JD relativistic}).\\

It has been showed in previous research the effect of external perturbations as noise and dissipation over the dynamics of some Hamiltonian systems (see Refs.~\cite{effectnoise, weaklynoise, noiseanddissipation}). For the sake of clarity, it is worth highlighting that the consideration of the relativistic framework on the system dynamics cannot be considered as an external perturbation like the noise or the dissipation, although the global properties of the system also change.\\


\section{Uncertainty dimension}\label{sec:fractal dimension}

\begin{figure}[htp]
\centering
\includegraphics[width=1.0\textwidth,clip]{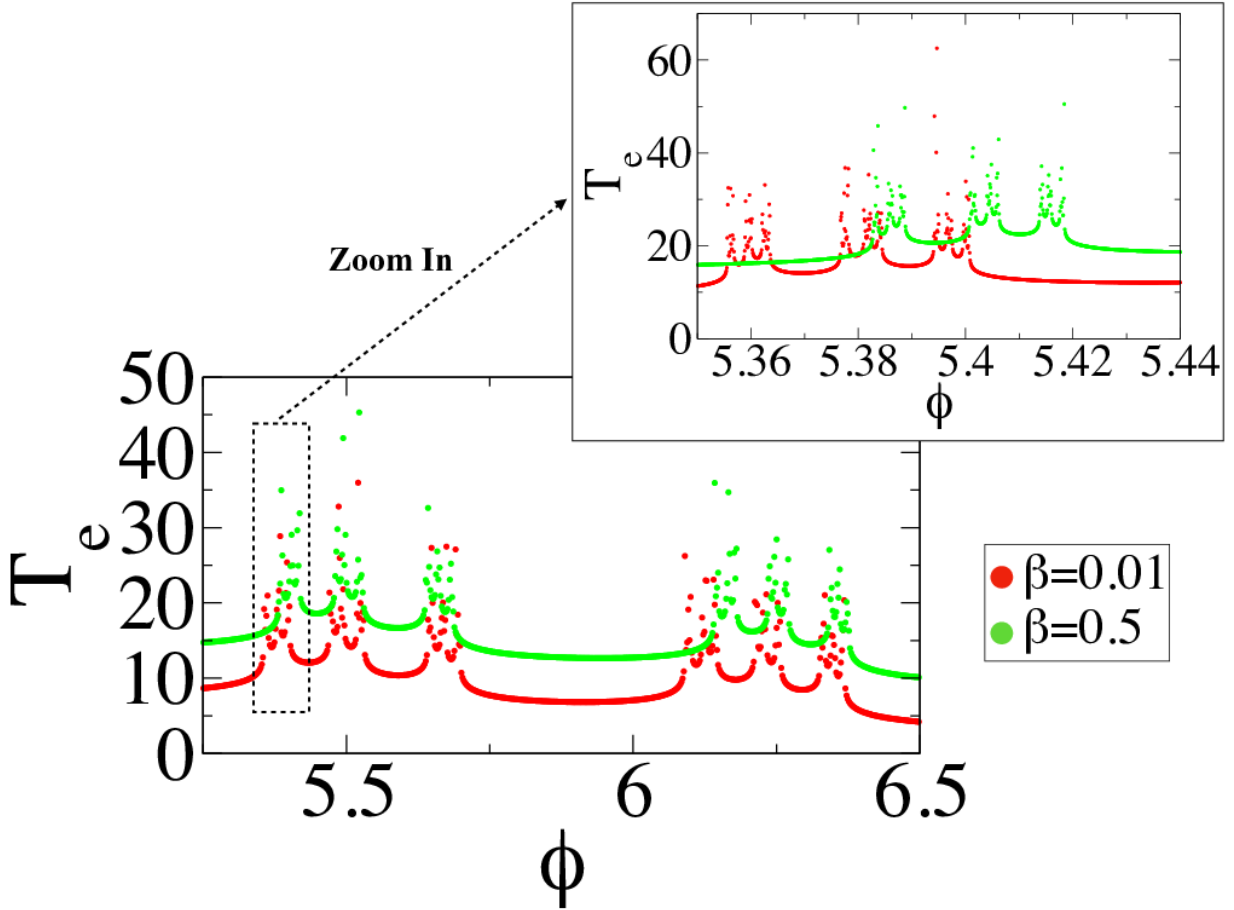}

\caption{(Color online) \textbf{Scattering Function.} Escape time $T_e$ \textit{vs.} the initial angle $\phi$ of $1,000$ particles shot into the scattering region from $(x,y)=(0,1)$ with initial velocity $v = 0.583$. Red (dark gray) dots represent the escape times for a relativistic system with $\beta = 0.01$, while green (gray) dots denote the escape time values for $\beta = 0.5$. The lower left panel shows the scattering function for a shooting angle $\phi\in[4.71,6.71]$. Likewise, the upper right panel is a zoom-in of the scattering function, taking a narrower initial angle range, $\phi\in[5.35,5.44]$. The scattering function contains some regions where the escape time of the particle varies smoothly with the shooting angle and, some others, where a slightly different initial condition in the shooting angle implies an abrupt change in the particle escape time.}
\label{Fig3}
\end{figure}

The \textit{scattering functions} are one the most fundamental footprints of any chaotic scattering systems. A scattering function relates an input variable of an incident particle with an output variable characterizing the trajectory of the particle, once the scattering occurs. These functions can be obtained empirically and, thanks to them, we can infer relevant information
about the system. In Fig.~\ref{Fig3}, we can see a typical scattering function: the average escape time of a test particle \textit{vs.} the initial shooting angle, for the relativistic H\'{e}non-Heiles system. Red (dark gray) dots are the values of the escape times for a relativistic system with $\beta = 0.01$, while green (gray) dots denote the escape times for $\beta = 0.5$. We use two panels to represent the scattering function. The lower left panel shows the scattering function for a shooting angle $\phi\in[4.71,6.71]$. The upper right panel is a magnification of the scattering function, varying the initial angle narrower, $\phi\in[5.35,5.44]$. In order to obtain both panels, we shoot in both cases $1,000$ particles from $(x,y) = (0,1)$ into the scattering region with an initial velocity $v = 0.583$. In Fig.~\ref{Fig3}, we can see that the scattering function contains some regions where the escape time of the particle varies smoothly with the shooting angle. However, there are some other fractal regions with singularities, where a slightly different initial condition in the shooting angle implies an abrupt change in the particle escape time.

The crucial point is that, because, as we have seen in Fig.~\ref{Fig3}, any small variation in the neighborhood of a singular input variable implies a huge variation in the  output variable, and furthermore, the range of variation of the output variable does not tend to zero despite the variation goes to zero. This type of behavior of the scattering function means that a small uncertainty in the input variable may make impossible any prediction about the value of the output variable. The fractal dimension $\alpha$ of the set of singular input variable values provides a quantitative characterization of the magnitude of such uncertainty. That is why the fractal dimension $\alpha$ is also called uncertainty dimension. As it was previously demonstrated, when the regime of the chaotic scattering system is hyperbolic, all the orbits are unstable and then $0 < \alpha < 1$. However, when the dynamics is nonhyperbolic, there are KAM islands in phase space and then $\alpha \approx 1$ \cite{Fractal dimension in nonhyperbolic Ott,Fractal dimension in hyperbolic}.\\

\begin{figure}[htp]
\centering
\includegraphics[width=0.65\textwidth,clip]{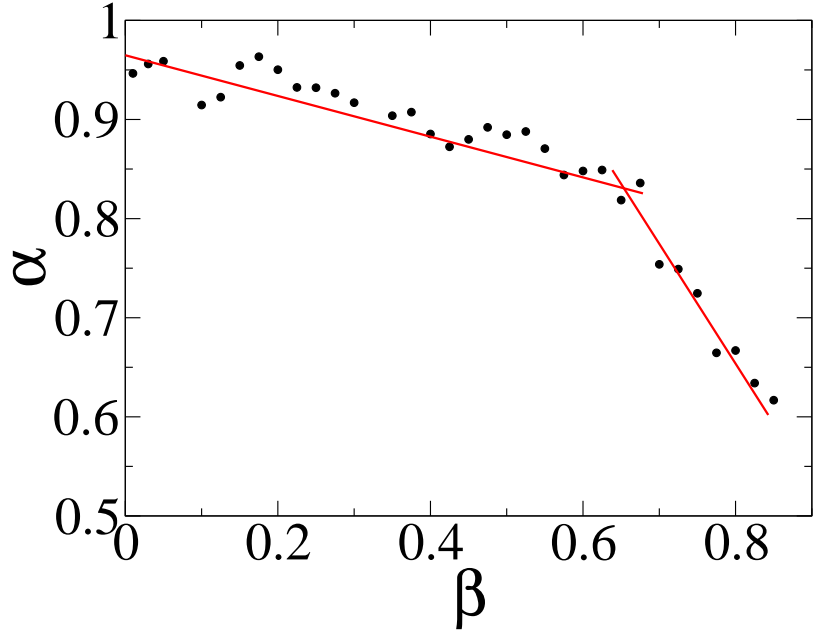}

\caption{(Color online) \textbf{Uncertainty dimension.} Evolution of the uncertainty dimension $\alpha$ in a scattering function defined on the horizontal initial line segment at $y_0 = 1$ with the variation of $\beta$. For many values of $\beta\in(0,1)$ we randomly launch $1,000$ test particles from the horizontal line segment passing through $y_0$. The particles are shot towards the scattering region with initial velocity $v = 0.583$. The results indicate that $\alpha\approx 1$ when $\beta\rightarrow 0$. Additionally, there is a linear decrease of $\alpha$ with any increment in $\beta$ up to a value $\beta\approx0.625$. At this point, there is a crossover behavior since, for values $\beta>0.625$, there is a steeper linear decrease of $\beta$.}
\label{Fig4}
\end{figure}

In this section we investigate the evolution of the uncertainty dimension, $\alpha$, in a typical scattering function as the parameter $\beta$ is varied. In order to compute $\alpha$, we use the uncertainty algorithm \cite{Grebogi}. We select a horizontal line segment defined by $y_0=1$ from which we shoot the test particles towards the scattering region with initial velocity $v = 0.583$. For a certain initial condition on the line segment, for instance $(x_0,y_0) = (0,1)$, we choose a perturbed initial condition $(x_0,y_0) = (x_0 + \varepsilon,1)$, where $\varepsilon$ is the amount of perturbation. Then we let both trajectories evolve according to Eq.~\ref{eq: relativistic eq motion}. We track the time they last in the scattering region, and by which exit they escape. In case that the two trajectories escape from the scattering region at the same time or throughout the same exit, then we consider that both trajectories are \textit{certain} with regard to the perturbation $\varepsilon$. Otherwise, we say both trajectories are \textit{uncertain}. Taking a large number of initial conditions for each value of $\varepsilon$, we can conclude that the fraction of uncertain initial conditions $f(\varepsilon)$ scales algebraically with $\varepsilon$ as $f(\varepsilon)\sim\varepsilon^{1-\alpha}$, or $f(\varepsilon)/\varepsilon\sim\varepsilon^{-\alpha}$. Therefore, $\alpha$ is the uncertainty dimension. When we repeat this process for different values of $\beta$, we obtain the evolution of the uncertainty dimension $\alpha$ with $\beta$. As we can see in Fig.~\ref{Fig4}, $\alpha\approx 1$ when $\beta\rightarrow 0$. Moreover, for $\beta\in[0,0.625)$, there is a linear decrease of $\alpha$ with any increment in $\beta$. There is a crossover behavior at $\beta\approx0.625$ such that for  $\beta>0.625$, the linear decrease of $\beta$ is steeper.

In order to provide a theoretical reasoning about the dependence of the uncertainty dimension $\alpha$ with the factor $\beta$, as shown in Fig.~\ref{Fig4}, we follow the approach explained in previous literature (see Ref.~\cite{Fractal dimension in dissipative Seoane}). Firstly, as an informative example, we consider a Cantor set with a Lebesgue measure equals zero and a fractal dimension equals $1$. To construct this set,
we proceed iteratively as follows. Iteration $1$: starting with the
closed interval $[0,1]$ of the real numbers, we remove the open
middle third interval. There are two remaining intervals of length
$1/3$ each. Iteration $2$: we remove the middle fourth interval
from the two remaining intervals and, therefore, we have four closed
intervals of length $1/9$. Iteration $3$: again, we take away the
fifth middle open interval from each four remaining intervals.
Iteration $n^{th}$: there are $N=2^n$ intervals, each of length
$\epsilon_n=2^{-n}[2/(n+2)]$. The total length of all the intervals
is $\epsilon_{n}N \sim n^{-1}$ and it goes to zero as $n$ goes to
infinity. For covering the set by intervals of size $\epsilon_n$,
the required number of intervals is $N(\epsilon) \sim \epsilon^{-1}
(\ln\epsilon^{-1})^{-1}$. On the other hand, the fractal dimension is
$\alpha=lim_{\epsilon\rightarrow0}[\ln(N(\epsilon))/\ln(\epsilon^{-1})]$,
which clearly yields $1$. We note that the exponent of the dependence $N(\epsilon) \sim 1/\epsilon^{\alpha}$ is the uncertainty dimension $\alpha$. The weaker logarithmic dependence does not have any influence on the determination of the dimension. However, the
logarithmic term is indeed the one that makes the Lebesgue measure equals zero since $\epsilon N \sim (\ln\epsilon^{-1})^{-1}$ tends to $0$ as $\epsilon \rightarrow 0$. In order to generalize this example, we may consider that in each stage we remove a fraction
$\eta_{n}=a/(n+b)$, where $a$ and $b$ are constants, from the middle of each of the $2^n$ remaining intervals. Then we find that

\begin{equation}\label{eq:Nepsilon}
N(\epsilon)\sim(1/\epsilon)[\ln(1/\epsilon)]^{-a}.
\end{equation}

According to Eq.~\ref{eq:Nepsilon}, the slope at any point of the curve $\ln N(\epsilon) vs \ln(1/\epsilon)$ is, by definition, $d \ln N(\epsilon)/d \ln(1/\epsilon)$, and it is always less than $1$ for small $\epsilon$, although it approaches
 $1$ logarithmically as $\epsilon \rightarrow 0$. Therefore, the result about the fractal dimension is still
 $\alpha=1$.\\

Coming back to the relativistic chaotic scattering analysis, now we can do a parallelism with the fractal dimension of Cantor-like structures. First, we note that chaotic scattering occurs due to a nonattracting chaotic set (\textit{i.e.}, a chaotic saddle) in phase space where the scattering interactions takes place~\cite{non-attracting chaotic set}. Moreover, both the stable and the unstable manifolds of the chaotic saddle are fractals~\cite{non-attracting chaotic set 2}.
Scattering particles are launched from a line segment straddling the stable manifold of the chaotic saddle outside the scattering region. The set of singularities is the set of intersections of the stable manifold and the line segment, and it can be effectively considered a Cantor-like set. There is an interval of input variables which leads to trajectories that remain in the scattering region for at least a duration of time $T_0$. By time $2T_0$, there is a fraction $\eta$ of these particles leaving the scattering region. In case that these particles are all located in the middle of the original interval, there are then two equal-length subsets of the input variable that lead to trajectories that remain in the scattering region for, at least, $2T_0$. Likewise, we may consider that a different fraction $\eta$ of incident particles, whose initial conditions were located in the middle of the first two subintervals that remains at time $2T_0$, are now leaving the scattering region by $3T_0$. There are then four particle subintervals that remains in the scattering region for at least $3T_0$. If we continue this iterative procedure, we can easily recognize the parallelism of the emerging fractal structure made of the incident particles which never escape, and a Cantor-like set of zero Lebesgue measure. The fractal dimension $\alpha$ of the Cantor set then is given by

\begin{equation}\label{eq:Dimensionscattering}
\alpha=\dfrac{\ln 2}{\ln[(1-\eta)/2]^{-1}}.
\end{equation}

In the nonhyperbolic regime, the decay law of the particles is algebraic and this implies that the fraction $\eta$ is not constant during the iterative process of construction of the Cantor set. At the $n^{th}$ stage (being $n$ large enough),
the fraction $\eta_{n}$ is approximately given by

\begin{equation}\label{eq:eta-nonhyperbolic}
\eta_{n} \approx -T_0P^{-1}dP/dt\approx z/n.
\end{equation}

This expression obviously yields a Cantor set with dimension
$\alpha=1$ when we substitute $\eta_{n} \approx z/n$ in Eq.~\ref{eq:Dimensionscattering}.
If we compare this result with the mathematical construction of the Cantor set as described in
Eq.~\ref{eq:Nepsilon}, then we realize that the exponent $z$ of the algebraic decay law corresponds
to the exponent $a$ of the Eq.~\ref{eq:Nepsilon}. On the other hand, in the case of the hyperbolic chaotic
scattering, the incident particles leave the scattering region exponentially.
The exponent of the decay law $\tau$ is related with the fraction $\eta$ as

\begin{equation}\label{eq:eta-hyperbolic}
\tau=T_0^{-1}\ln(1-\eta)^{-1}.
\end{equation}

Now, we are in the position to complete the reasoning behind the behavior of the uncertainty dimension $\alpha$ with the relativistic parameter $\beta$. As we pointed out in Sec.~\ref{sec:Model Description}, the relation between the parameter
$\tau$ of the hyperbolic regime and $\beta$ follows a quadratic relation. Then, according to
Eqs.~\ref{eq:Dimensionscattering} and \ref{eq:eta-hyperbolic} we may find that

\begin{equation}\label{eq:Dimensionscatteringexponential}
\alpha=\dfrac{\ln2}{ \ln2+T_0(\Gamma_0 + \Gamma_1\beta +
\Gamma_2\beta^2)}.
\end{equation}

According to Eq.~\ref{eq:Dimensionscatteringexponential}, the fractal dimension $\alpha$ is always less that $1$ and it decreases while $\beta$ increases. In the limit, as $\beta \rightarrow 1$, the quadratic relation between the particles decay rate $\tau$ and $\beta$ is no longer valid, since the kinetic energy of the system grows up to $+\infty$. In that case, as $\beta \rightarrow 1$, then $\tau \rightarrow +\infty$ and, therefore, $\alpha \rightarrow 0$. Likewise, the exponent $z$ of the algebraic decay law and the parameter $\beta$ are both related in a quadratic manner too. Then if we take into consideration Eqs.~\ref{eq:Dimensionscattering} and \ref{eq:eta-nonhyperbolic} we obtain

\begin{equation}\label{eq:Dimensionscatteringalgebraic}
\alpha=\dfrac{\ln2}{\ln(\frac{2}{1-(Z_0 + Z_1\beta +
Z_2\beta^2)/n})}.
\end{equation}

As $\beta \rightarrow 0$, we have $\alpha \rightarrow 1$ for large $n$, and, moreover, we find that $d\alpha/d\beta=0$ since we recover the Newtonian system.
When $\beta$ increases, then $\alpha$ decreases. For large values of $n$, the value of
$1-(Z_0 + Z_1\beta + Z_2\beta^2)/n$ is always larger than $0$ and smaller than $1$ because the maximun value of $z$ is $z\approx1.5$, as we noted in Sec.~\ref{sec:Model Description}.
Despite the KAM destruction, the transition from the algebraic regime to the hyperbolic
one is not very abrupt.
This is the reason why the uncertainty dimension $\alpha$ decreases smoothly with $\beta$ as shown in
Fig.~\ref{Fig4} up to $\beta \approx 0.625$. When we reach this value, the hyperbolic regime is clear and it yields a steeper change in $\alpha$ \textit{vs.} $\beta$.


\section{Exit basins description} \label{sec:Escape Basins}

We define \textit{exit basin} as the set of initial conditions whose trajectories converges to an specified exit~\cite{escape basins definition}. Likewise, an initial condition is a \textit{boundary point} of a basin $B$ if every open neighborhood of $y$ has a nonempty intersection with basin $B$ and at least one other basin. The $boundary$ of a basin is the set of all boundary points of that basin. The basin boundary could be a smooth curve, but in chaotic systems, the boundaries are usually fractal. In this case, since the phase space resolution is finite in any real physical situation, those fractal structures impose an extreme dependence on the initial conditions, which obstructs the prediction of the system final state. For that reason, the understanding of the exit basin topology is crucial to foresee the final fate of the system. In this section, we will provide a qualitative description of how the exit basins of the H\'{e}non-Heiles system evolve while the ratio $\beta$ is increased.\\

\begin{figure}[htp]
\centering
\includegraphics[width=0.4\textwidth,clip]{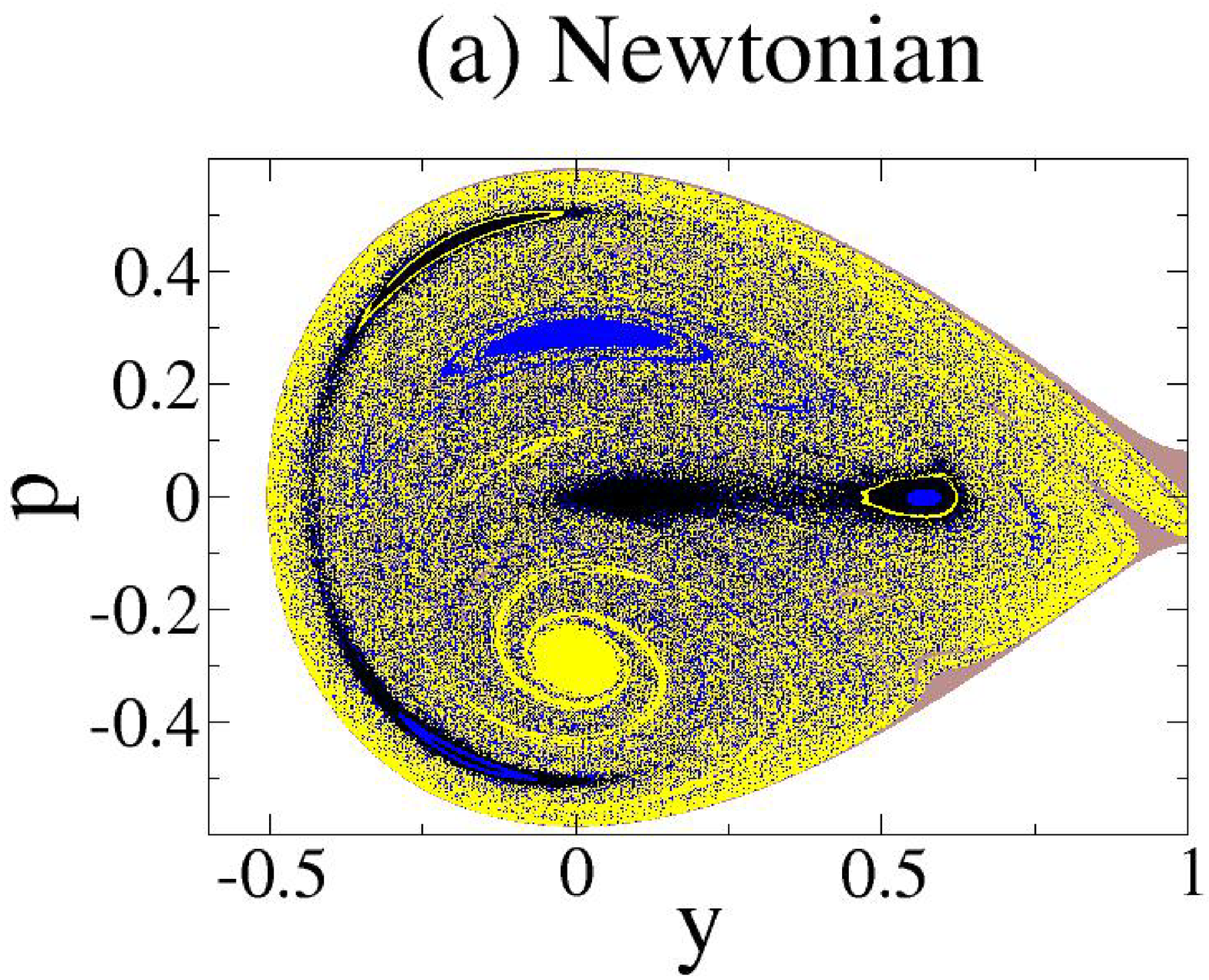}\hspace{.25 cm}
\includegraphics[width=0.4\textwidth,clip]{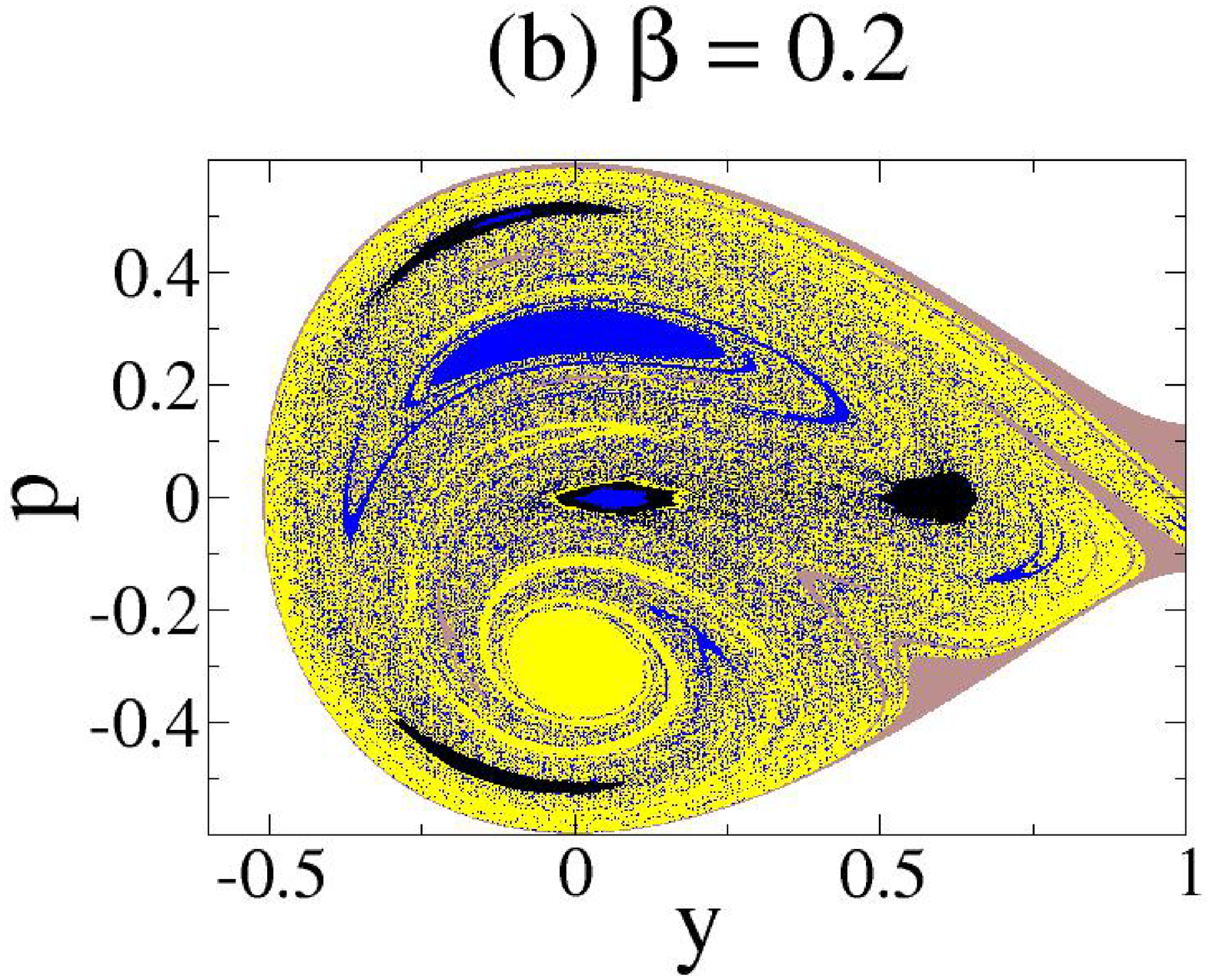}\hspace{.25 cm}
\includegraphics[width=0.4\textwidth,clip]{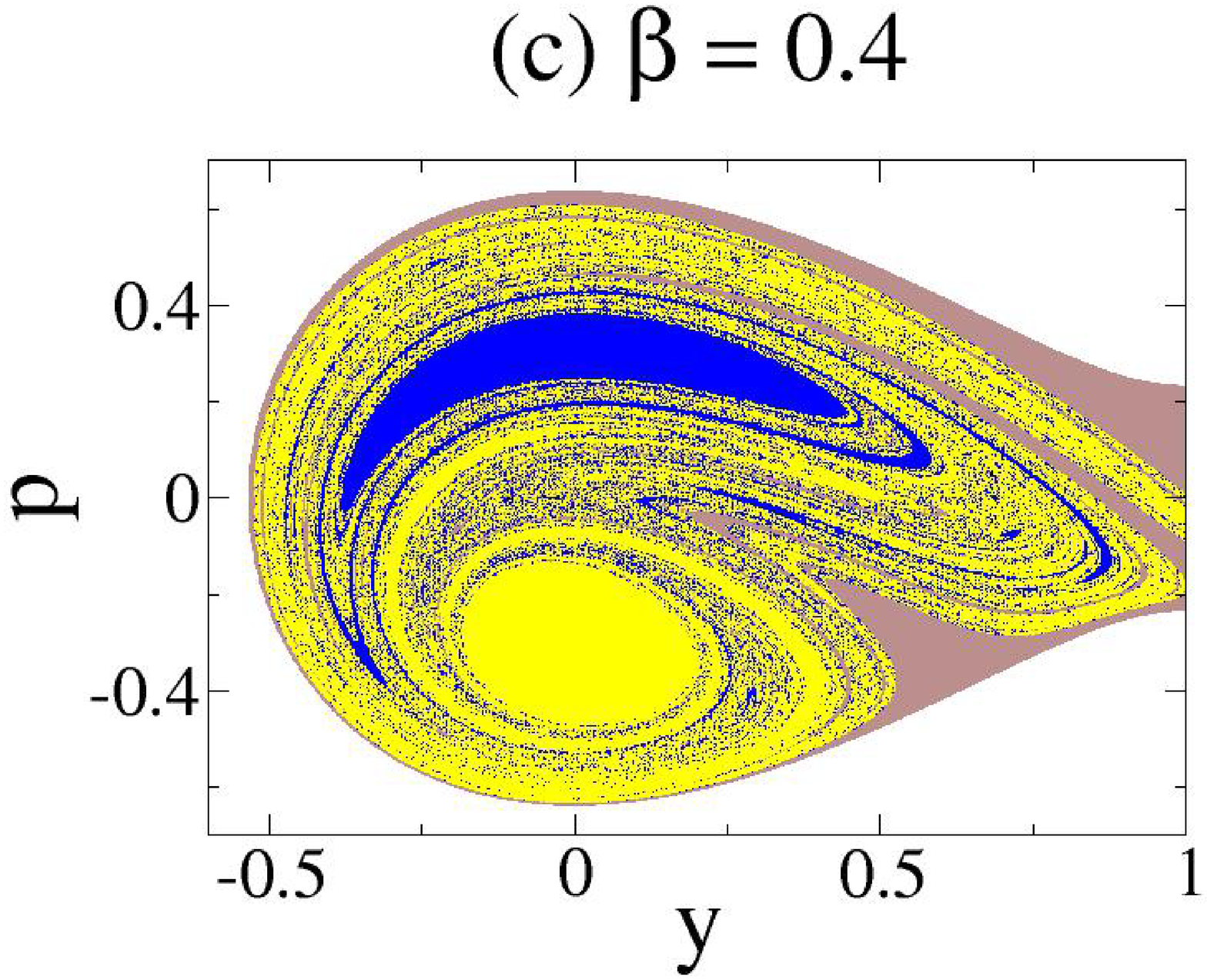}\hspace{.25 cm}
\includegraphics[width=0.4\textwidth,clip]{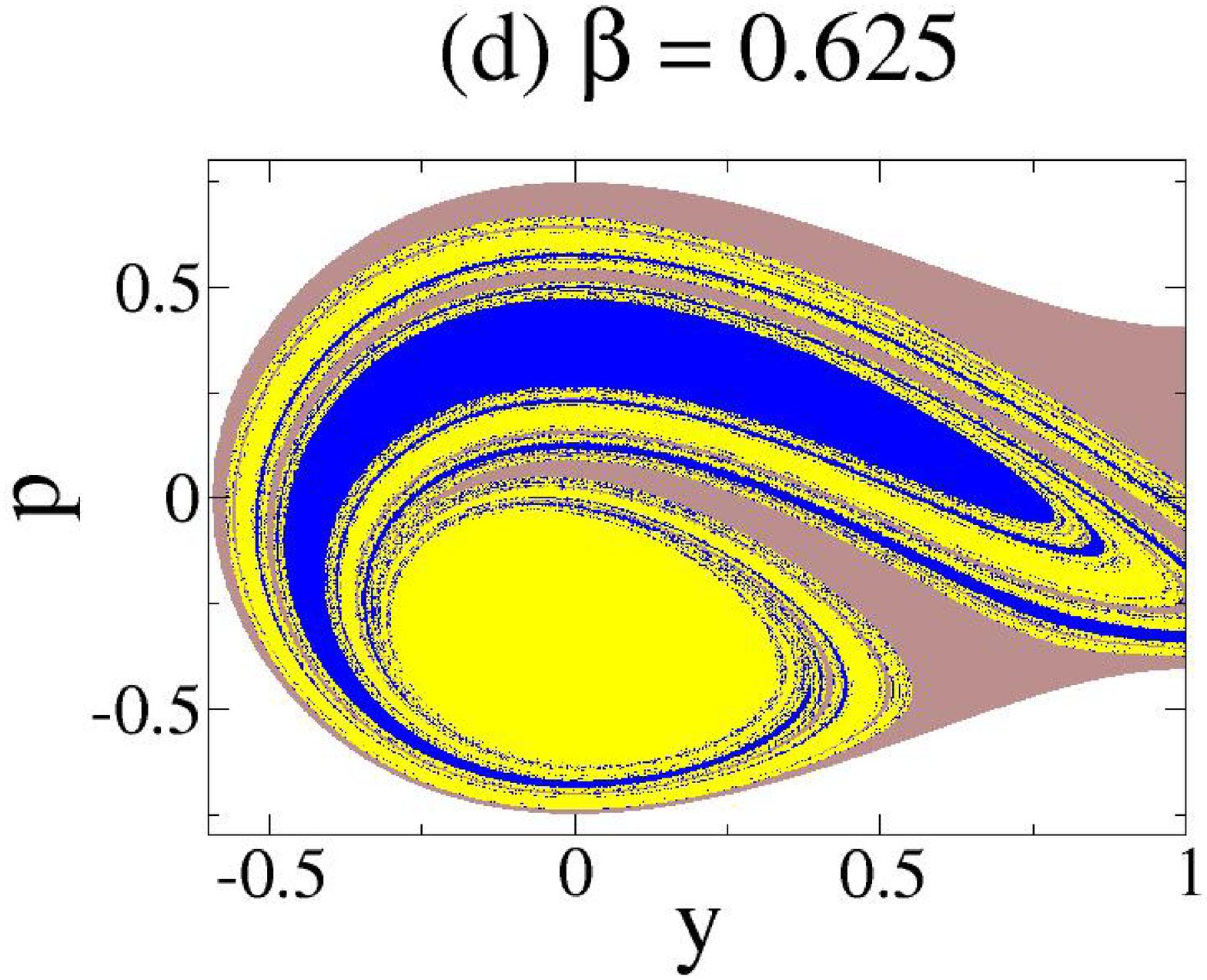}\hspace{.25 cm}
\includegraphics[width=0.4\textwidth,clip]{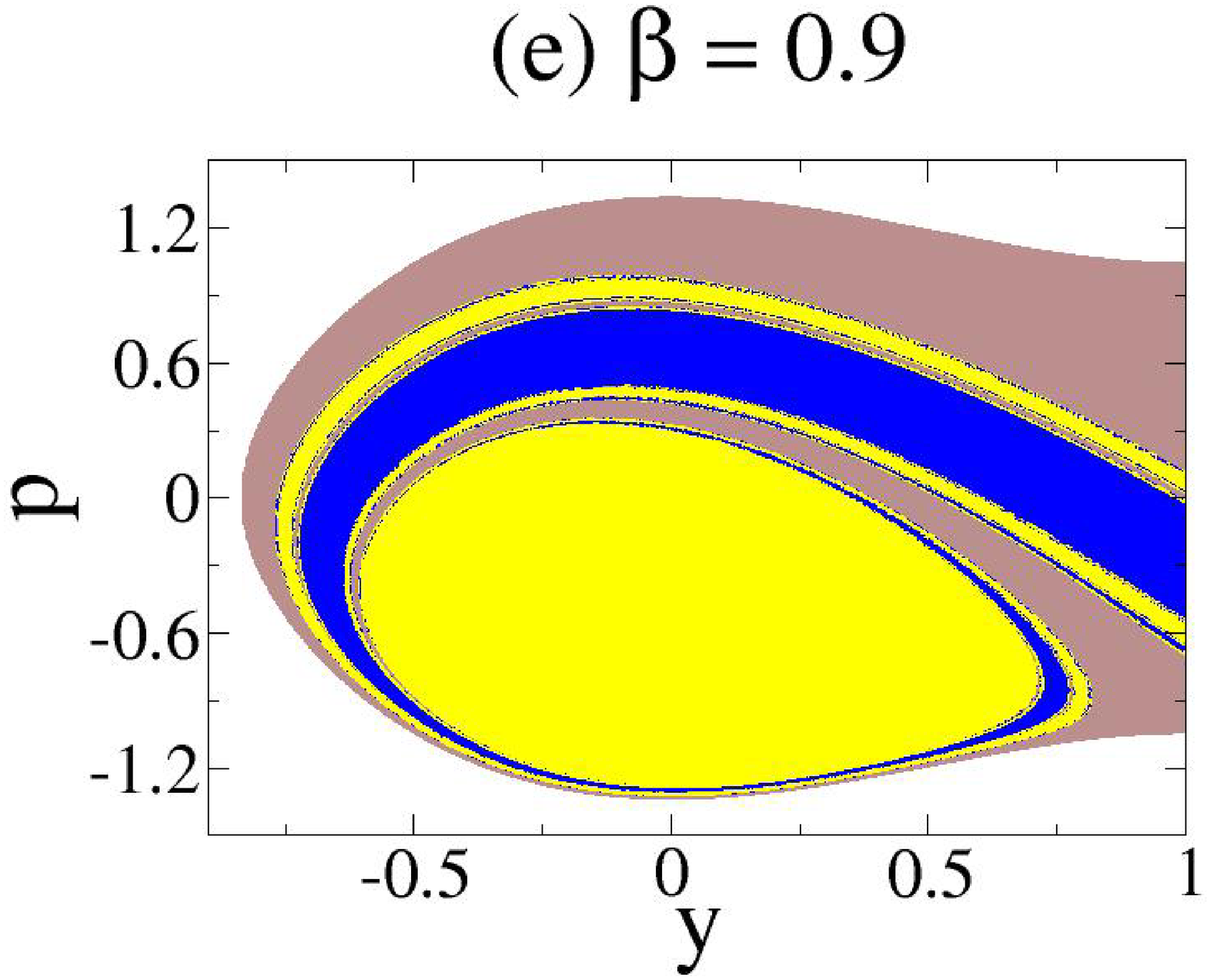}

\caption{(Color online) \textbf{Evolution of the exit basins of the H\'{e}non-Heiles system for different values of $\beta$.} The sets of brown (gray), blue (dark gray) and yellow (light gray) dots denote initial conditions resulting in trajectories that escape through Exits $1$, $2$ and $3$ (see Fig.~\ref{Fig1}), respectively. The black regions denote the KAM islands and, generally speaking, the black dots are the initial conditions which do not escape. (a) Newtonian case: the exit basins are quite mixed throughout the phase space and the KAM islands can be easily recognized as the big black regions inside phase space. (b) $\beta = 0.2$: the exit basins are still fairly mixed. The regions corresponding to exit basins are larger than in the Newtonian case. (c) $\beta = 0.4$: exit basin regions are larger. The exit basin boundaries are fractal. The KAM islands are destroyed. (d) $\beta = 0.625$: the exit basins are not mixed anymore. The boundaries are fractal. (e) $\beta = 0.9$: the boundaries are smoother, and the exit basins occupy a larger region of phase space.}
\label{Fig5}
\end{figure}

We use Poincar\'{e} section surfaces $(q,y)$ at $t=0$ and $x = 0$ to represent the exit basins. To carry out our simulations, we shoot $1,000,000$ particles from $x = 0$ and $y\in[-1,1]$, with initial angles $\phi\in[0,\pi]$. Then, we follow each trajectory and we register by which exit the particles have escaped from the scattering region. If a particle leaves the scatterer by Exit $1$, then we color the initial condition, $(q_0,y_0)$, in brown (gray). Likewise, we color the initial condition in blue (dark gray) if the particle escapes by Exit $2$ and, in case it leaves the scatterer by Exit $3$, we color it in yellow (light gray). When a particle remains in the scattering region after $t_{max}$, then we color its initial condition in black. We have run different simulations to plot the exit basins of the H\'{e}non-Heiles system for a wide range of parameters $\beta$, as shown in Fig.~\ref{Fig5}.

In Figs.~\ref{Fig5}(a-e), we can see the evolution of the exit basins of the H\'{e}non-Heiles system while the parameter $\beta$ is increased. In Fig.~\ref{Fig5}(a), we represent the Newtonian case. It corresponds to $E_N = 0.17$, which is fairly close to the escaping energy threshold value $E_e = 1/6$. At this energy, the exit basins are quite mixed throughout the phase space and there are many initial conditions (black dots) which do not escape from the scattering region. Likewise, the KAM islands can be easily recognized as the black areas inside the phase space. In Fig.~\ref{Fig5}(b), we can see the relativistic effects of the Lorentz corrections for $\beta = 0.2$. The exit basins are still quite mixed although now there are larger regions where we can see compact exit basins. In Fig.~\ref{Fig5}(c), we represent the exit basins of the system for $\beta = 0.4$. The exit basins are clearly located in regions and their boundaries are fractal. There are still some portions of the phase space where the basins corresponding to Exits 1, 2 and 3 are mixed. However, the KAM islands are destroyed and there are just a few trajectories remaining in the scattering region after $T_{max}$ (colored in black). Figure~\ref{Fig5}(d) represents the exit basins of the system for $\beta = 0.625$. There are no regions where the exit basins are mixed. The boundaries are fractal.  Lastly, in Fig.~\ref{Fig5}(e), we show the Hen\'{o}n-Heiles exit basins for $\beta=0.9$. The exit basins are smoothly spread on the phase space and the fractality of the boundaries has decreased.\\

\begin{figure}[htp]
\centering
\includegraphics[width=0.65\textwidth,clip]{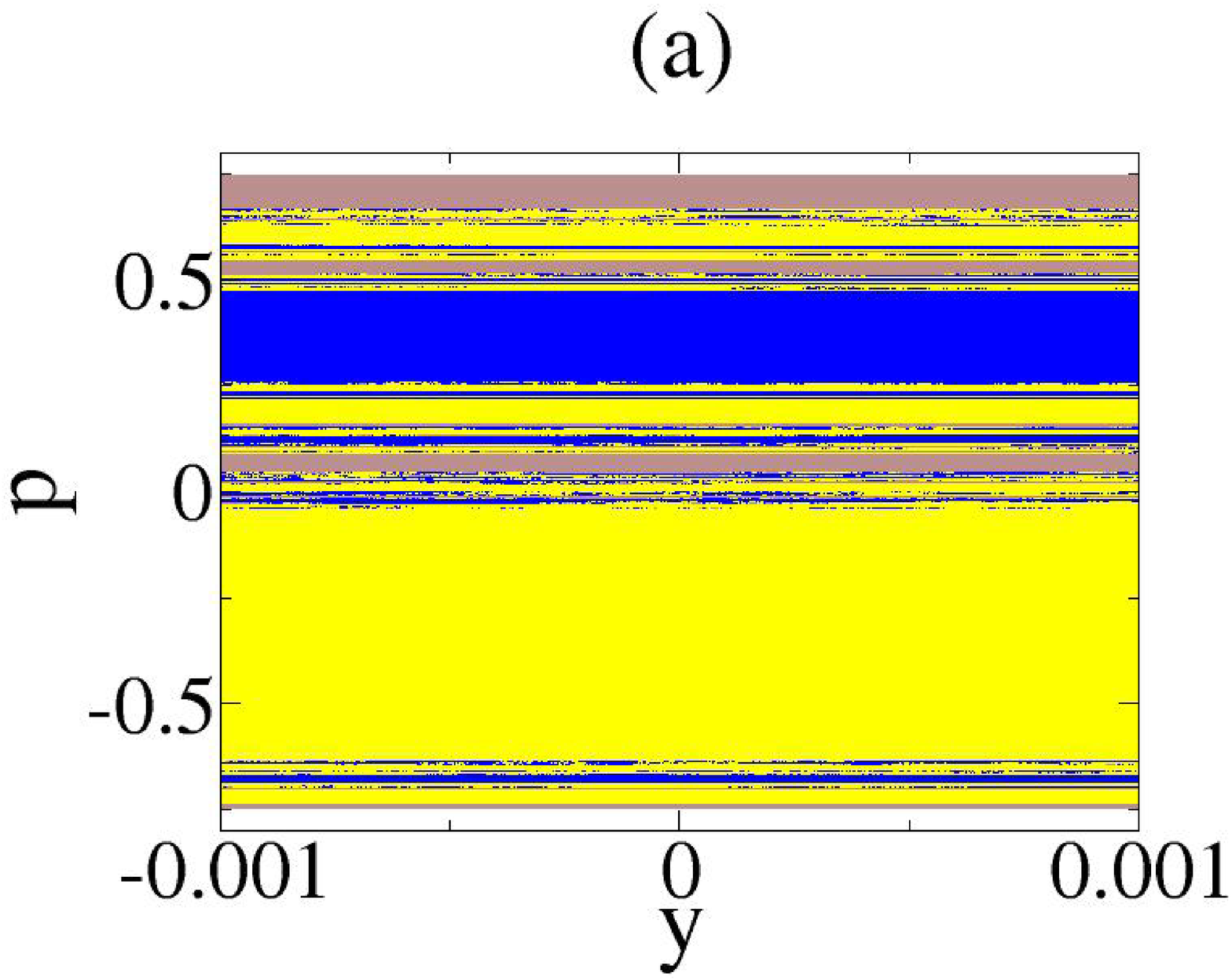}\hspace{.25 cm}
\includegraphics[width=0.65\textwidth,clip]{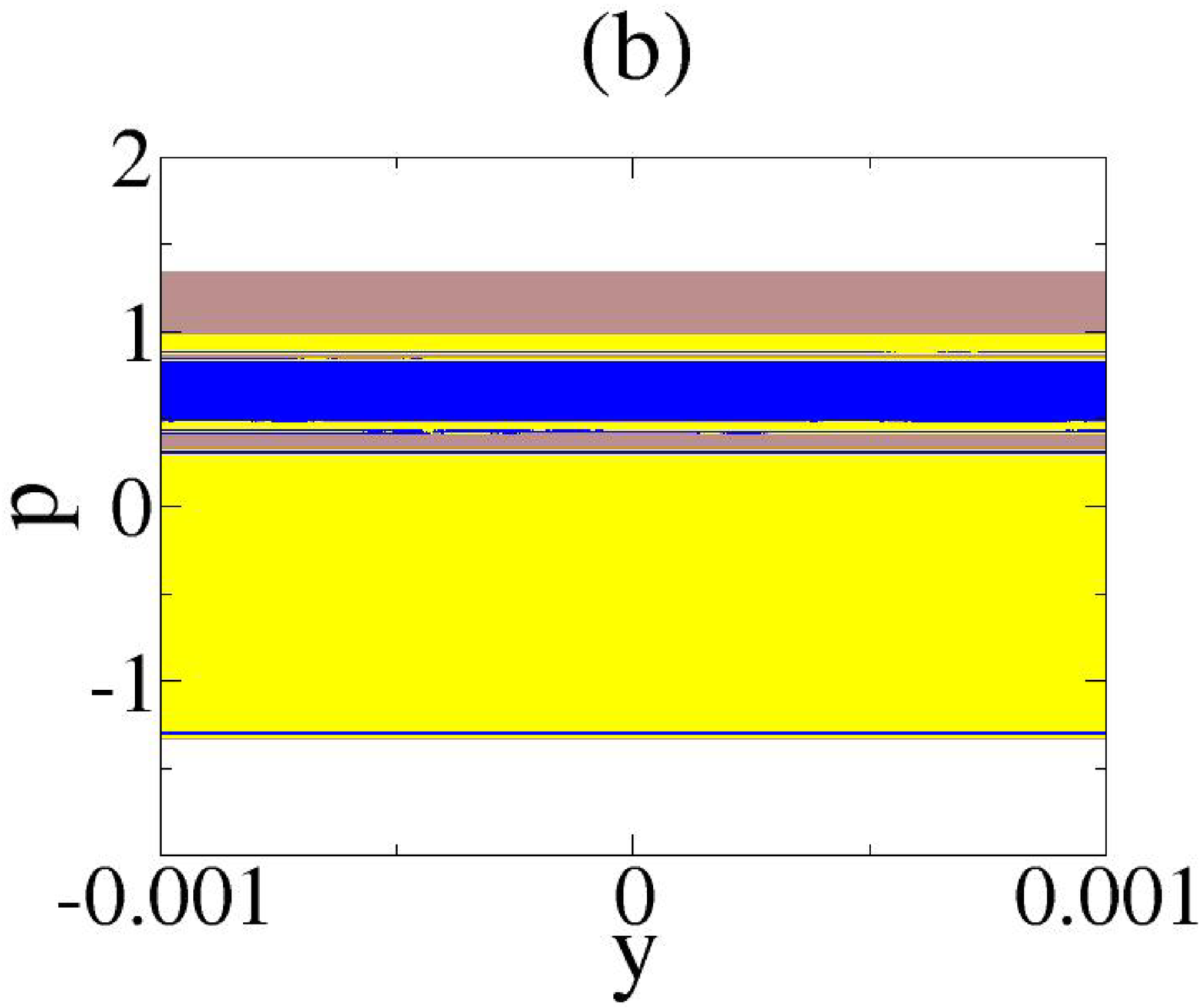}

\caption{(Color online) \textbf{Zoom-in of the exit basins for $\beta=0.625$ and $\beta=0.9$.} The sets of brown (gray), blue (dark gray) and yellow (light gray) dots denote initial conditions resulting in trajectories that escape through Exits 1, 2 and 3 (see Fig.~\ref{Fig1}), respectively. In (a) we can see that the boundary points seem to be surrounded by points from other basins. Nonetheless, in (b) we can find boundary points which are surrounded only by points of just two basins. The boundary located at $p=1$ is smooth and it divides just two basins, the one corresponding to Exit $1$ from the corresponding to the Exit $3$.}
\label{Fig6}
\end{figure}

Many dynamical system of interest, with two or more coexisting
attractors (or escapes), exhibit a singular topological property in
their basin boundaries that is called the \textit{property of Wada}.
If for every boundary point $u_b$ of a certain basin, we can find an
infinitely tiny open neighborhood centered in $u_b$ that contains
points from the rest of the basins, we can say that this boundary
has the property of Wada~\cite{Wada}. A logical consequence of this
definition is that a Wada basin boundary is the same boundary for
all the basins. The property of Wada is a very interesting
characteristic because the fate of any dynamical system is harder to
predict since we cannot foresee a priori by which of the exits any
initial condition close to the boundaries is going to escape. In
that case, the degree of unpredictability of the destinations can be
more severe than the case where there are just fractal basins with
only two potential destinations associated. In the energy regime
that we are considering, the H\'{e}non-Heiles system exhibits three
symmetric exits to escape from the scattering region, giving rise to
three qualitatively distinct scattering destinations. This allows
Wada basin boundaries to occur. Each exit of the system has its own
associated exit basin. Previous research described some algorithms
to obtain the numerical verification of the Wada property in
dynamical systems ~\cite{Helena, Jacobo, Seoane Wada, SanjuanWada}.
In this paper, we resort to the appearance of the exit basin
boundaries to give visual indications about the persistence of the
Wada property as the parameter $\beta$ is varied. For values of
$\beta \leq 0.625$, we have observed that the boundary points of any
exit basin magnification seem to be surrounded by points from the
three basins. However, when $\beta > 0.625$, we have seen that the
boundary points are exclusively surrounded by points belonging just
to two basins. In order to give visual examples, we show
Figs.~\ref{Fig6}(a) and (b). Both are detailed analysis of the
H\'{e}non-Heiles exit basins for $\beta=0.625$ and $\beta=0.9$,
respectively, when we perform the computations on a tiny portion of
the exit basins, in $y\in[-0.001,0.001]$. Therefore,
Fig.~\ref{Fig6}(a) is a zoom-in of the Fig.~\ref{Fig5}(d) and
Fig.~\ref{Fig6}(b) is a zoom-in of the Fig.~\ref{Fig5}(e). For
$\beta>0.625$, the exit basin representations are similar to the one
described in Fig.~\ref{Fig6}(b), where we can observe that, for
example, the boundary located at $p=1$ is smooth. Moreover, this
boundary divides only two basins, the one corresponding to Exit $1$
(brown--gray--) from the one associated to Exit $3$ (yellow--light
gray--). This is a visual indication that the Wada property might
not be observable in the relativistic H\'{e}non-Heiles system for
$\beta>0.625$; at least, at the numerical scale we have performed
the calculations. In that sense, we may suppose that the
unpredictability associated to the final destination of the
trajectories for values of $\beta \leq 0.625$ is higher.

As we can see, the consideration of the special relativity have qualitative implications on the exit basin topology of the system, even for low values of $\beta$. In the following sections we will provide more insights about this statement from a quantitative point of view.


\section{Basin entropy}\label{sec:Basin Entropy}
The \textit{basin entropy} is a novel tool developed to
quantitatively describe the exit basin topology of any dynamical
system ~\cite{Alvar}. The main idea behind the concept of basin
entropy is that the continuous phase space of the system can be
considered a discrete grid due to the finite resolution of any
experimental or numerical procedure to determine any point in phase
space. In fact, this unavoidable scaling error indeed does induce
wrong predictions in chaotic systems even when they are completely
deterministic. Therefore, the basin entropy helps to quantify to
what extent one system is chaotic according to the topology of its
phase space. In particular, considering the exit basins of the
H\'{e}non-Heiles sytem as the ones shown in Fig.~\ref{Fig5}, we can
easily create a discrete grid if we assume a finite precision
$\delta$ in the determination of the initial conditions and we cover
the phase space with boxes of size $\delta$. This way, every piece
of the grid is surrounded by other pieces, and we may define a
\textit{ball} around a piece as the pieces sharing some side with
it. The method to calculate the basin entropy considers that, the
ball is a random variable, being the potential results of that
variable the different exit basins. Taking into account that the
pieces inside the ball are independent and applying the Gibbs
entropy concept, the basin entropy $S_b$ is defined as

\begin{equation}\label{eq:basin entropy}
S_b = \sum_{k=1}^{k_{max}}\dfrac{N_{k}^{0}}{N^{0}}\delta^{\alpha_{k}}\log(m_{k}),
\end{equation}
where $k$ is the label for the different exit basin boundaries,
$m_{k}$ is the number of exit basins contained in a certain ball and
$\alpha_{k}$ is the uncertainty dimension of the boundary $k$ as
defined in Sec.~\ref{sec:fractal dimension}. The ratio
$\dfrac{N_{k}^{0}}{N^{0}}$ is a term related with the portion of the
discretized phase space occupied by the boundaries, that is, the
number of pieces lying in the boundaries divided by the total number
of pieces in the grid. Therefore, there are three sources that
increase the basin entropy: (a) $\dfrac{N_{k}^{0}}{N^{0}}$, that is,
the larger portion of the phase space occupied by the boundaries,
the higher $S_b$; (b) the uncertainty dimension term
$\delta^{\alpha_{k}}$, related to the fractality of the boundaries;
(c) $\log(m_{k})$, which is a term related to the number of
different exit basins $m_{k}$. In the case that the basins exhibit
the property of Wada, then there is just one boundary that separates
all the basins. In this case, the term $\log(m_k)$ is maximun and
$S_b$ is increased because all the possible exits are present in
every boundary box. As we have shown in Sec.~\ref{sec:Escape
Basins}, this may be the case for the relativistic H\'{e}non-Heiles
system for $\beta \leq 0.625$.\\

\begin{figure}[htp]
\centering
\includegraphics[width=0.65\textwidth,clip]{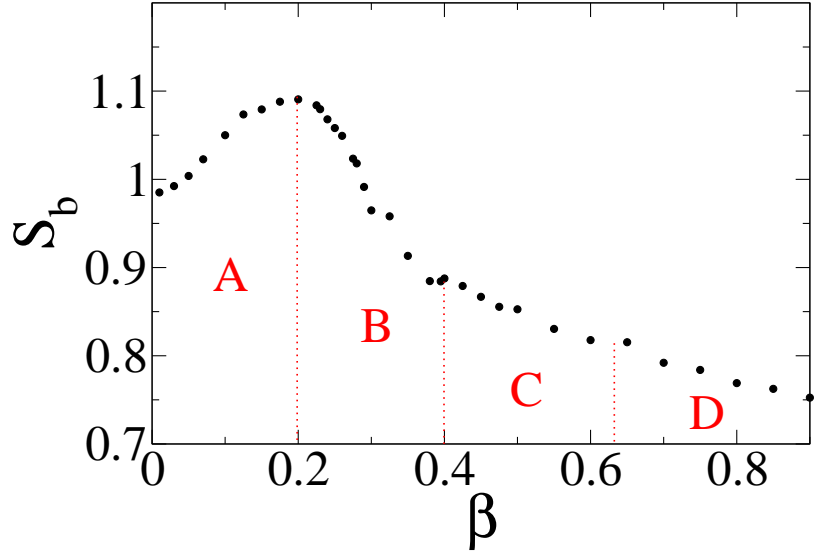}
\caption{(Color online) \textbf{Basin entropy.} The figure shows the evolution of the basin entropy $S_b$ of the relativistic H\'{e}non-Heiles system with $\beta$. There are represented 4 regions: (A) $\beta\in(0,0.2]$, increase of $S_b$ up to $\beta\approx0.2$; (B) $\beta\in[0,0.4]$, steep decrease of $S_b$ until $\beta\approx0.4$; (C) $\beta\in(0.4,0.625]$ and (D) $\beta\in(0.625,0.9)$, both regions show a smoother decrease of $S_b$. The behavior of $S_b$ in region (A) is explained because of the reduction of both the KAM islands and the regions where the basins are mixed. The global effect is that $\dfrac{N_{1}^{0}}{N^{0}}$ is increased. In contrast, in region (B) the areas where the basins are mixed are negligible and the KAM islands are rapidly tinnier. These effects cause a relevant reduction of $\dfrac{N_{1}^{0}}{N^{0}}$. In regions (C) and (D) the exit basin sets are progressively larger and smoother (as was shown in Fig.~\ref{Fig5}).}
\label{Fig7}
\end{figure}

In Fig.~\ref{Fig7}, we can see the evolution of the basin entropy $S_b$ of the H\'{e}non-Heiles system with $\beta$. We can distinguish 4 regions: (A) $\beta\in(0,0.2]$, increasing of $S_b$ up to $\beta\approx0.2$; (B) $\beta\in[0,0.4]$, steep decrease of $S_b$ until $\beta\approx0.4$; (C) $\beta\in(0.4,0.625]$ and (D) $\beta\in(0.625,0.9)$, smoother decrease of $S_b$. As was shown in Sec.~\ref{sec:fractal dimension}, the uncertainty dimension $\alpha$ is a monotonically decreasing function, so the increase of $S_b$ in the region (A) can only be explained because of a higher increase of $\dfrac{N_{1}^{0}}{N^{0}}$. In region (A), when $\beta$ is increased, the zones where the basins are mixed are indeed reduced. However, the KAM islands are reduced. These effects can be seen in the exit basin evolution from  Fig.~\ref{Fig5}(a) to Fig.~\ref{Fig5}(b). Moreover there are more pieces in the grid of the discretized phase space occupied by the boundaries. In the region (B) there is an important decrease of $S_b$ because of the reduction of $\dfrac{N_{1}^{0}}{N^{0}}$. In this region, while $\beta$ is increased, the areas of the phase space where the basins are mixed are negligible and the KAM islands are progressively losing relevance in phase space. At $\beta\approx0.4$ there is an inflection point, just when the KAM islands are destroyed. There are fewer pieces of the grid occupied by the boundaries. In the region (C), $\beta\in(0.4,0.625]$, the exit basin areas are larger and they grow as $\beta$ increases, while the fractality of the boundaries decreases. This is exactly what was found in Fig.~\ref{Fig4}, where the uncertainty dimension $\alpha$ decreases abruptly from $\beta \approx 0.625$ on. In region (D), the fractality of the boundaries is reduced and there have been some visual indications about the disappearance of the Wada basins, as described in Sec.~\ref{sec:Escape Basins}.

As we have seen along this work, the exit basin topology of the relativistic H\'{e}non-Heiles varies with $\beta$, even for low velocities. From this point of view, the properties of the system that depend on the phase space topology may vary too. Although the H\'{e}non-Heiles Hamiltonian was initially developed to model the motion of stars around an axisymetrical galaxy, we considered that the results described in the present work may be extrapolated to many other real phenomena occurred in Nature. Examples of systems that can be described by the H\'{e}non-Heiles Hamiltonian are, for instance, the planar three-body system, buckled beams, and some stationary plasma systems~\cite{Bastos}. One relevant topic of special interest that is also modeled by the  H\'{e}non-Heiles Hamiltonian is the dynamics of charged particles in a magnetic dipole field (wich is also called the St\"{o}rmer problem). Since a long time, scientists have studied it in the context of the northern lights and cosmic radiation; and it models how a charged particle moves in the magnetic field of the Earth. The analysis of this system leads to the conclusion that charged particles are trapped into the Earth magnetosphere or escape to infinity, and the trapping region is bounded by a torus-like surface, the Van Allen inner radiation belt. In the trapping region, the motion of the charged particles can be periodic, quasi-period or chaotic ~\cite{Stormer}. According to the results we have presented in this paper, if we want to study global properties of the St\"{o}rmer problem, we should consider the relativistic effects because those properties depend on the exit basin topology. In those cases, we may expect that the uncertainty associated to the prediction of the final state of the particles varies as the parameter $\beta$ increases.


\section{Conclusions} \label{chap: conclusions}

In the past years there have been relevant progress in understanding the effects of the special relativity in the field of chaotic scattering. Most of the previous research has been focused on studying the discrepancies between the Newtonian and the relativistic approaches over the trajectories of single particles. More recently, there has been demonstrated that some global properties of chaotic scattering systems, as the average escape time and the particles decay law, do depend on the effect of the Lorentz transformations, even for low velocities.\\

In the present work we have focused our attention in describing
different characteristics of the exit basin topology as the
uncertainty dimension, the Wada property and the basin entropy of
the relativistic H\'{e}non-Heiles system. We have found that the
Lorentz corrections modify these quantities. We have shown in
Fig.~\ref{Fig3} the evolution of the uncertainty dimension,
$\alpha$, in a typical scattering function as the parameter $\beta$
is varied. We found that $\alpha$ decreases almost as $\beta$
increases up to a certain value of the parameter $\beta \approx
0.625$, when a crossover phenomenon occurs and $\alpha$ decreases
abruptly. This takes place due to the transition from an algebraic
particles decay law to a exponential decay law. We have carefully
explained how the uncertainty dimension $\alpha$ varies with
$\beta$. We have also described in a qualitative manner the
evolution of the exit basin topology with the parameter $\beta$ (see
Fig.~\ref{Fig5}). Moreover, according to the numerical resolution we
have used in the computations of the exit basins, we have found some
visual indications that the Wada basin boundaries may not be present
for $\beta > 0.625$. Lastly, we have used the concept of basin
entropy to quantify the evolution of the exit basins with any
variations of the parameter $\beta$ (see Fig.~\ref{Fig7}). We have
resorted to the three sources of change of the basin entropy to
explain the obtained results. All our results point out that the
uncertainty in the prediction of the final fate of the system
depends on the considered value of $\beta$, and this relation is not
linear. We have observed that $S_b$ increases as $\beta$ grows in
the interval $\beta\in(0,0.2]$, while $S_b$ decreases repidly for
$\beta\in(0.4,0.625]$. Therefore, if we want to make accurate
predictions about the final state of any chaotic scattering system,
we think that the relativistic corrections should always be
considered, regardless the energy of the system. That would be the
case, for example, of the global properties of charged particles
moving through a magnetic dipole-field as the one modeled by the
St\"{o}rmer problem ~\cite{Bastos, Stormer}. We consider that our
results can be useful for a better understanding of relativistic
chaotic scattering systems.

\section*{ACKNOWLEDGMENTS}
This work was supported by the Spanish State Research Agency (AEI) and the European Regional Development Fund (FEDER) under Project No. FIS2016-76883-P. MAFS acknowledges the jointly sponsored financial support by the Fulbright Program and the Spanish Ministry of Education (Program No. FMECD-ST-2016).

\end{document}